\documentclass[twocolumn,prb,amsmath,amssymb,showpacs,superscriptaddress]{revtex4}

\usepackage{hyperref}
\usepackage{graphicx}

\renewcommand{\Re}{\mathop\mathrm{Re}\nolimits}
\renewcommand{\Im}{\mathop\mathrm{Im}\nolimits}

 \DeclareMathOperator{\Tr}{\mathrm{Tr}}
 \DeclareMathOperator{\STr}{\mathrm{STr}}

\begin{document}

 \title{Effective spin-flip scattering in diffusive superconducting proximity systems\\ with magnetic disorder}

 \author{D.~A.~Ivanov}
 \affiliation{Institute of Theoretical Physics, Ecole Polytechnique F\'ed\'erale de Lausanne (EPFL), CH-1015 Lausanne,
              Switzerland}

 \author{Ya.~V.~Fominov}
 \affiliation{L.~D.~Landau Institute for Theoretical Physics RAS, 119334 Moscow, Russia}

 \author{M.~A.~Skvortsov}
 \affiliation{L.~D.~Landau Institute for Theoretical Physics RAS, 119334 Moscow, Russia}

 \author{P.~M.~Ostrovsky}
 \affiliation{Institut f\"ur Nanotechnologie, Forschungszentrum Karlsruhe, 76021 Karlsruhe, Germany}
 \affiliation{L.~D.~Landau Institute for Theoretical Physics RAS, 119334 Moscow, Russia}

\date{5 October 2009}

\begin{abstract}
We revisit the problem of diffusive proximity systems involving
superconductors and normal metals (or ferromagnets) with magnetic
disorder. On the length scales much larger than its correlation
length, the effect of sufficiently weak magnetic disorder may be
incorporated as a local spin-flip term in the Usadel equations. We
derive this spin-flip term in the general case of a three-dimensional
disordered Zeeman-type field with an arbitrary correlation
length. Three different regimes may be distinguished: pointlike
impurities (the correlation length is shorter than the Fermi
wavelength), medium-range disorder (the correlation length between the
Fermi wavelength and the mean free path), and long-range disorder (the
correlation length longer than the mean free path). We discuss the
relations between these three regimes by using the three overlapping approaches:
the Usadel equations, the non-linear sigma model, and the diagrammatic
expansion. The expressions for the spin-flip rate agree with the
existing results obtained in less general situations.

\centerline{\textbf{Published as: Phys.\ Rev.~B \textbf{80}, 134501 (2009)}}
\end{abstract}






\maketitle

\tableofcontents

\section{Introduction}

In conventional superconductors (S), pairing occurs between electrons
with opposite spins, and thus, the coexistence of superconductivity and
magnetism may lead to a variety of interesting effects in
superconducting and proximity structures. Examples of such effects are
gapless superconductivity,\cite{AG} triplet proximity correlations
(see Ref.~\onlinecite{BVE_review} for a review), and Josephson
$\pi$-junctions (see Refs.~\onlinecite{Buzdin,Golubov} for a
review). In a ferromagnet (F) with a uniform exchange field, the
theory of anomalous correlations may be constructed by taking into
account the splitting of electronic bands of opposite
spins.\cite{Buzdin} The situation becomes more complicated if the
magnetic structure is inhomogeneous. In this case, if the
ferromagnetic structure is non-collinear, the triplet component of
anomalous correlations needs to be taken into
consideration.\cite{BVE_review} The case of inhomogeneous magnetic
structure is also interesting from the practical point of view, since
many experimental studies of hybrid superconductor-ferromagnet (SF) systems, in particular,
Josephson $\pi$-junctions, reveal a strong spin-flip
scattering.\cite{Sellier,Aprili,Ryazanov1,Ryazanov2} A theoretical
analysis of superconducting correlations in the presence of an inhomogeneous magnetic structure is
complicated, and many of the existing studies are limited to considering
either pointlike magnetic impurities\cite{AG,Gorkov-Rusinov,Fulde-Maki,Buzdin,Houzet,Efetov,LS}
or specific domain
geometries.\cite{BVE_review,Buzdin_av1,Buzdin_av2,BBK,Blanter,Eschrig,FVE,Radovic,CTI,BN,Asano,KKG,HV,Linder,Silaev}

In a recent work by two of the present authors, the spin-flip term (in the same form as for magnetic impurities\cite{AG}) has been derived in the model of non-collinear magnetic disorder correlated on length scales much larger than the elastic mean free path.\cite{IF} The resulting expression for the spin-flip rate agrees with the known one obtained for the collinear inhomogeneous magnetization under assumption of periodic magnetic structure.\cite{Buzdin_av1,Buzdin_av2}

In the present paper, we revisit the problem of the magnetic disorder in the more general situation of the non-collinear disorder with arbitrary correlation length. We consider a superconducting or proximity-type system with potential impurities and an inhomogeneous Zeeman field. The potential impurities are supposed to be sufficiently strong to bring the electronic motion to the diffusive regime. On top of this diffusive motion, the electrons experience splitting from the inhomogeneous Zeeman field, which is assumed to be random and Gaussian with an arbitrary pair correlation. We further assume that this magnetic disorder is much weaker than the potential one, in terms of the characteristic scattering rates.

Then three different regimes can be distinguished: the short-range magnetic disorder (or, equivalently, pointlike impurities, with the correlation length shorter than the Fermi wavelength), the medium-range disorder (with the correlation length between the Fermi wavelength and the elastic mean free path), and the long-range disorder (the correlation length longer than the elastic mean free path). The short-range case has been solved in Ref.~\onlinecite{AG}, the medium- and long-range regimes have been treated in Refs.~\onlinecite{Buzdin_av1,Buzdin_av2} for the collinear periodic case, and the long-range non-collinear case was studied in Ref.~\onlinecite{IF}. We extend those results to the general non-collinear case and remove some of the technical assumptions made in Ref.~\onlinecite{IF}.

\begin{figure}[t]
 \includegraphics[width=0.95\hsize]{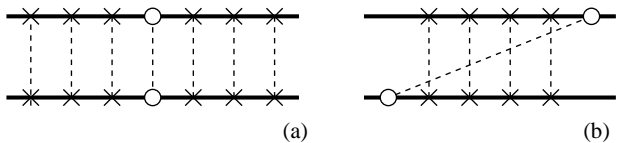}
\caption{Two types of spin-flip scattering: (a)~local; (b)~nonlocal. The diagrams show the way the magnetic scattering is included in the ``cooperon'' propagator (see Secs.~\ref{section:sigma-model} and~\ref{section:diagram} for details). Thick solid lines represent the electron Green functions, thin dashed lines denote averaging over Gaussian disorder. The crosses and the open circles represent the potential and magnetic disorders, respectively.}
 \label{fig:diagr}
\end{figure}

We use three methods for our analysis: the non-linear sigma model, the
Usadel equations, and the diagrammatic technique. While the calculations
in these three methods are somewhat parallel to each other, we find
it instructive to present those various approaches, in order to illustrate
the correspondence between the methods and to clarify the physical
meaning and the applicability conditions of the results. As we shall see
below, the spin-flip term in the short-range and medium-range regimes
corresponds to inserting one magnetic-impurity rung into the cooperon
ladder [Fig.~\ref{fig:diagr}(a)], while the long-range regime
corresponds to the magnetic line crossing many cooperon rungs
[Fig.~\ref{fig:diagr}(b)].

The paper is organized as follows. In Sec.~\ref{section:results}, we
present the main results of the paper: the form of the spin-flip term
in the sigma-model and Usadel description, as well as the expressions
for the spin-flip scattering rate in the three regimes, including the
crossovers between those regimes. In the following sections, we
present details of the calculations. In
Sec.~\ref{section:sigma-model}, we employ the sigma model to study all
cases of the magnetic-disorder correlation length except for the
crossover between the medium- and long-range regimes. In
Sec.~\ref{section:usadel}, we demonstrate how the long-range regime
can be treated in the language of the Usadel equations. In
Sec.~\ref{section:diagram}, we consider all the cases with the help of
the diagrammatic technique.  Finally, in
Sec.~\ref{section:conclusions}, we present our conclusions.

\section{Main results}
\label{section:results}

\subsection{Parameters of the problem}

We consider a diffusive motion of electrons in a finite piece of magnetic metal (i.e., a metal with exchange interaction inside of it)
of linear size $L$. The Thouless energy scale (the inverse diffusion time) of this region is
$E_\mathrm{Th}=D/L^2$, where $D$ is the diffusion constant (we put $\hbar=1$). Below we study the anomalous
correlations in this magnetic metal induced either by a small (see conditions below)
superconducting order parameter $\Delta$ or by an electric contact
with a superconductor (proximity-induced superconductivity,
in which case the order parameter $\Delta$ is put to zero in the magnetic metal).
The electrons and the (Andreev-reflected) holes are
considered at a finite energy $E$ (relative to the Fermi level).
We further consider an exchange field in the magnetic metal
$\mathbf H(\mathbf r) = \mathbf h(\mathbf r)+\delta \mathbf h(\mathbf r)$
containing a small (possibly zero) smooth (varying
on the length scale $L$) part ${\mathbf h}$
and the disorder component $\delta \mathbf h$
(the typical scale of the disorder part will be further denoted as $\delta h$).
The magnetic disorder is assumed to be correlated on a short length scale $a$,
which defines the scale of ``the Thouless energy of magnetic
inhomogeneities'' $E_a=D/a^2$. We assume a
Gaussian ensemble for $\delta \mathbf h$, with the pair
correlation function
\begin{equation} \label{disorder}
\bigl< \delta h_i (\mathbf r) \delta h_j (\mathbf r') \bigr>
= F_{ij} \left( |\mathbf r-\mathbf r'| \right).
\end{equation}
The typical order of magnitude for $ F_{ij} (r)$ is then $(\delta
h)^2$, and the typical support is of order $a$.

For the energy scales in the magnetic metal, the following condition is assumed:\cite{T}
\begin{equation} \label{hierarchy-general}
E_\mathrm{Th}, E, \Delta, h, \Gamma_\mathrm{sf} \ll E_a, \tau^{-1} \, ,
\end{equation}
where $\Gamma_\mathrm{sf}$ is the resulting effective spin-flip rate, and $\tau$ is the mean free time due to potential scattering.
The physical meaning of this condition is that the length scales
associated with both potential and magnetic disorder (the right-hand
side of the inequality) are much shorter than the length scales
involved in the Usadel equations (the left-hand side of the
inequality).

\subsection{Spin-flip term in the sigma model}

The sigma-model action has the form
\begin{equation} \label{action-full}
\mathcal S[Q] = \mathcal S_0 + \mathcal S_\mathrm{sf}\, ,
\end{equation}
where the usual sigma-model action is
\begin{multline} \label{action-0}
\mathcal S_0  = \pi\nu \int d^3 \mathbf r\; \STr\Big\{ \frac D4 (\nabla Q)^2 \\
+ \left( i E \hat\tau_3 - \hat\Delta - i \mathbf h (\mathbf r) \hat\tau_3 \hat{\boldsymbol\sigma} \right) Q \Big\}
\end{multline}
(we use the standard notations of the sigma-model technique, see
Sec.~\ref{section:sigma-model} for definitions),
and the spin-flip term is
\begin{equation} \label{action-sf}
\mathcal S_\mathrm{sf} = -\frac{\pi\nu}2 \int d^3 \mathbf r\; \Gamma_\mathrm{sf}^{ij} \STr\left(  \hat\tau_3 \hat\sigma_i Q \hat\tau_3 \hat\sigma_j Q \right) \, ,
\end{equation}
where $\Gamma_\mathrm{sf}^{ij}$ is a symmetric matrix of the spin-flip scattering rates.\cite{spin-flip-matrix} Here and below, we use the convention of
summing over repeating indices.

\subsection{Spin-flip term in the Usadel equations}

The spin-flip term in the Usadel equations can be obtained by varying the action (\ref{action-full}) with respect to the $Q$ matrix. Denoting the saddle-point value of the $Q$ matrix as the matrix Green function $\check g$, we write the resulting equation as
\begin{multline} \label{Usadel_symm}
D\nabla \left( \check g \nabla \check g \right) +\left[ i E \hat\tau_3 \hat\sigma_0 - \hat\Delta \hat\sigma_0 - i\hat\tau_3 (\mathbf h \hat{\boldsymbol\sigma}), \check g \right] \\
-\Gamma_\mathrm{sf}^{ij} \left[ \hat\tau_3 \hat\sigma_i \check g \hat\tau_3 \hat\sigma_j, \check g \right] = 0.
\end{multline}
The non-linear constraint $\check g^2=1$ can be resolved by a parametrization. In
the case when the phase of the anomalous Green function is fixed (in which case
the Usadel equations contain four parameters), we may use the parametrization of
Ref.~\onlinecite{IF} in terms of the angle $\theta$ and the vector $\mathbf M$:
\begin{equation} \label{Q-parameterization}
\check g = M_0 \hat\sigma_0(\hat\tau_3 \cos \theta + \hat\tau_1 \sin \theta) + i \mathbf M \hat{\boldsymbol\sigma}
(\hat\tau_3 \sin \theta - \hat\tau_1 \cos \theta)
\end{equation}
with the constraint
\begin{equation} \label{normalization}
M_0^2 - \mathbf M^2 = 1.
\end{equation}
A generalization to the case of a varying phase is straightforward, in which case four more parameters need to be added, corresponding to the terms containing $\hat\tau_2$ in the $\check g$ matrix.

Substituting the parametrization (\ref{Q-parameterization}) into Eq.\ (\ref{Usadel_symm}), we obtain the triplet Usadel equations with the spin-flip terms:
\begin{align}
\frac D2 & \nabla^2 \theta + M_0 \left( i E \sin\theta + \Delta \cos\theta \right) - ( \mathbf{h M} ) \cos\theta
\notag \\
& -\left( \Gamma_\mathrm{sf}^\mathrm{tot} + 2\mathbf M^T \hat\Gamma_\mathrm{sf} \mathbf M \right) \sin 2\theta =0 \, ,
\label{Us1} \\[10pt]
\frac D2 & \left( \mathbf M \nabla^2 M_0 - M_0 \nabla^2 \mathbf M \right) - \mathbf M ( i E \cos\theta - \Delta \sin\theta )
\notag \\
& - M_0 \mathbf h \sin\theta + 2 M_0 \hat\Gamma_\mathrm{sf} \mathbf M  \cos 2\theta =0 \, ,
\label{Us2}
\end{align}
where $\hat\Gamma_\mathrm{sf}$ is the symmetric $3\times 3$ matrix of $\Gamma_\mathrm{sf}^{ij}$, and  $\Gamma_\mathrm{sf}^\mathrm{tot} = \Tr \hat\Gamma_\mathrm{sf}$. This set of equations generalizes the conventional spin-flip term\cite{Kopnin} to the case of the triplet Usadel equations.\cite{spin-flip-matrix}

\subsection{Spin-flip scattering rate}

\subsubsection{Regime of short-range correlations}
\label{subsection-short}

In the regime of $a \ll k_F^{-1}$, the spin-flip rates are given by\cite{AG}
\begin{equation} \label{sf-1}
\Gamma_\mathrm{sf}^{ij} = \pi\nu \int d^3 \mathbf r\; F_{ij}(r) \sim \nu \left( \delta h \right)^2 a^3\, .
\end{equation}

\subsubsection{Regime of medium-range correlations}
\label{subsection-medium}

In the regime of $k_F^{-1} \ll a \ll l$, the spin-flip rates are given by
\begin{equation} \label{sf-2}
\Gamma_\mathrm{sf}^{ij} =  \pi \nu \int d^3 \mathbf r\; F_{ij}(r) \frac{1}{2 (k_F r)^2} \sim \nu \left( \delta h \right)^2 a k_F^{-2}\, .
\end{equation}

The exact formula interpolating between the two regimes (\ref{sf-1}) and (\ref{sf-2}) has the form
\begin{equation} \label{sf-1-2}
\Gamma_\mathrm{sf}^{ij} =  \pi \nu \int d^3 \mathbf r\; F_{ij}(r) \frac{\sin^2 (k_F r)}{(k_F r)^2}\, .
\end{equation}

\subsubsection{Regime of long-range correlations}
\label{subsection-long}

In the regime of $l \ll a$, the spin-flip rates are given by
\begin{equation} \label{sf-3}
\Gamma_\mathrm{sf}^{ij} =  \frac 1D \int d^3 \mathbf r\; F_{ij}(r) \frac 1{4\pi r} \sim \nu \left( \delta h \right)^2 \frac{a^2}l k_F^{-2}\, .
\end{equation}

The exact formula interpolating between the two regimes (\ref{sf-2}) and (\ref{sf-3}) has the form
\begin{equation} \label{sf-2-3}
\Gamma_\mathrm{sf}^{ij} =  \frac{l^2}{3D} \int \frac{d^3 \mathbf q}{(2\pi)^3}\; F_{ij}(q) \; \frac{\arctan (ql)}{ql-\arctan(ql)}\, .
\end{equation}

Note that the diffusion constant is related to the density of states by $D=v_F l/3 = k_F^2 l/(6\pi^2\nu)$.

The results (\ref{sf-3}) and (\ref{sf-2-3}) have been derived for collinear periodic magnetic structures in Refs.~\onlinecite{Buzdin_av1,Buzdin_av2}. The result (\ref{sf-3}) in the non-collinear isotropic case has also been found in Ref.~\onlinecite{IF}.

\section{Sigma-model derivation}
\label{section:sigma-model}

In the sigma-model description, if we assume that the random exchange field $\delta\mathbf h$ is sufficiently weak, then it can be included in the sigma-model action perturbatively. For the delta-correlated (short-range) magnetic disorder, this procedure is well known,\cite{Efetov,LS} and our consideration generalizes it to arbitrary correlation lengths.

In our derivation, we find that there are two different contributions arising from the magnetic disorder. By expanding the action in the disordered field to the second order,
\begin{equation} \label{S0-S1-S2}
\mathcal S = \mathcal S_0 + \mathcal S_1 + \mathcal S_2\, ,
\end{equation}
we find the two spin-flip-type contributions:
\begin{equation} \label{two-contributions}
\mathcal S_\mathrm{loc} = \bigl< \mathcal S_2 \bigr> \quad \mathrm{and} \quad \mathcal S_\mathrm{nonloc} = -\frac 12 \bigl< (\mathcal S_1)^2 \bigr> \, .
\end{equation}
We shall see below, that the first (``local'') contribution dominates
in the case of short-range or medium-range disorder and gives the
spin-flip rate (\ref{sf-1-2}), while the second (``nonlocal'')
contribution becomes dominant in the regime of long-range disorder and
gives the spin-flip rate (\ref{sf-3}). In performing the averaging for
the nonlocal contribution, one needs to take into account fluctuations
around the replica-symmetric (or supersymmetric) saddle point. On
inspection, the local and nonlocal spin-flip contributions correspond
to the processes depicted in Fig.~\ref{fig:diagr} (left and right
panel, respectively). Interpolating between these two regimes goes beyond
the scope of the sigma-model derivation in this Section, but it is done
in Sec.~\ref{section:diagram} in the diagrammatic language.

\subsection{Sigma-model action}

To derive the sigma model for a disordered superconducting (or proximity) system, we follow the usual procedure.\cite{Efetov,AST} The potential disorder is assumed to be Gaussian and $\delta$-correlated,
\begin{equation} \label{U}
\bigl< U(\mathbf r) U(\mathbf r') \bigr>_U = \frac{\delta(\mathbf r -\mathbf r')}{2\pi\nu\tau}
\end{equation}
[here $\nu = m k_F / (2 \pi^2)$ is the density of states at the Fermi level per one spin projection], while the magnetic disorder is also taken to be Gaussian, but with an arbitrary correlation length, see Eq.\ (\ref{disorder}).

As usual in the derivation of the sigma model, we assume the ``dirty limit'', i.e., that
\begin{equation} \label{dirty-1}
E_\mathrm{Th}, E, \Delta, h \ll \tau^{-1}\, .
\end{equation}
In addition, we need that the effective spin-flip scattering (whose rate we derive below) is much weaker than the potential scattering,
\begin{equation} \label{dirty-2}
\Gamma_\mathrm{sf} \ll \tau^{-1}\, ,
\end{equation}
so that we can treat it as a perturbation on top of the diffusive sigma model defined by the potential disorder.

The derivation of the sigma-model action starts with the
partition function for excitations with energy $E$ in the Bogolyubov -- de Gennes Hamiltonian including also the exchange field:\cite{LS}
\begin{align}
 \mathcal Z = \int D\Psi^* & D\Psi \; e^{-\mathcal S},
\\
 \mathcal S [\Psi] =
 -i\int d^3 \mathbf r\; \Psi^+ \biggl[
 & E - \mathbf H \hat{\boldsymbol\sigma} - \hat\tau_3 \Bigl( \xi + U(\mathbf r) \Bigr)
\notag \\
 & - \hat\tau_2 \Re\Delta - \hat\tau_1 \Im\Delta \biggr] \Psi,
 \label{act2} \\
\xi = -\frac 1{2m} \frac{\partial^2}{\partial \mathbf r^2} & -\mu.
\end{align}
Here, $\Psi^*$ and $\Psi$ are 4-component (in the product of the Nambu-Gor'kov and spin spaces) fermionic vector fields containing Grassmann anticommuting elements. For brevity, we do not write a small imaginary part $i0$ that should be added to the energy $E$. The Pauli matrices $\hat\sigma_i$ act on the spin of electrons while $\hat\tau_i$ act in the Nambu-Gor'kov space.

Now we introduce replicas\cite{Finkelstein,E} (or supersymmetry\cite{Efetov}) for averaging over the potential disorder, thus extending the $\Psi$ vector.
As a result of averaging with the disorder correlator (\ref{U}), we obtain
\begin{multline} \label{act4}
 \mathcal S[\Psi]
  = -i \int d^3 \mathbf r\; \Psi^{+} \biggl[
      E - \mathbf H
\hat{\boldsymbol\sigma} - \hat\tau_3 \xi
      - \hat\tau_2 \Re\Delta \\
      - \hat\tau_1 \Im\Delta
    \biggr] \Psi + \frac 1{4\pi\nu\tau}
\int d^3 \mathbf r\, \left( \Psi^{+}
    \hat\tau_3 \Psi \right)^2.
\end{multline}

Next, we perform the Hubbard--Stratonovich transformation with the
help of a matrix field $Q(\mathbf r)$. The $Q$ field is a matrix in
the product of the Nambu-Gor'kov and spin spaces and also in the space
of replicas (or in the Fermi-Bose superspace). As a result of the
Hubbard--Stratonovich transformation, the action becomes quadratic,
and the Gaussian integration over $\Psi^*$ and $\Psi$ yields
\begin{align}
 \mathcal S[Q]
  = \frac{\pi\nu}{4\tau}\int d^3\mathbf r\; &\STr Q^2 \notag \\
    -\int d^3\mathbf r\; \STr \ln & \left[
      \xi -
        \hat\tau_3 \left( E - \mathbf H \hat{\boldsymbol\sigma} \right)
        -i \hat\Delta - \frac{i Q}{2\tau}
    \right], \label{action}
 \\
 & \hat\Delta
  = \begin{pmatrix}
      0 & \Delta \\
      \Delta^* & 0 \\
    \end{pmatrix} .
\end{align}
The ``$\STr$'' operation is the complete trace. In particular, it includes the trace over the replica indices or the supertrace, depending on the version of the sigma model (replica or supersymmetric).

In general, the $Q$ matrix of a sigma model for a superconducting
system would contain all the diffusive soft modes: diffusons and
cooperons. This would be achieved by an additional doubling of the
fermionic fields,\cite{Efetov,AST,LS} and hence of the $Q$ matrix, in
the retarded-advanced space. The doubled $Q$ matrix would be subject
to an additional constraint reflecting the symmetry of the Bogolyubov
-- de Gennes Hamiltonian.  Note that our sigma model does not involve
this doubling and accordingly includes only the cooperon modes but
not the diffusons. As a result, our calculations are valid only
at the saddle-point level (Usadel equations), but cannot reproduce
weak-localization corrections involving diffuson modes. An interesting
question of the effect of the inhomogeneous magnetic disorder on the
weak-localization corrections requires a more detailed analysis and
is postponed until further studies.
In this paper, since we are interested in the effect of the magnetic disorder on the saddle point, we use the reduced version of the sigma model.

In the quasiclassical regime, where the Fermi energy is the largest energy scale, $E_F \tau \gg 1$, the $Q$ matrix is restricted to the manifold\cite{Efetov,AST}
\begin{equation}
Q^2=1\, .
\end{equation}
Furthermore, using the dirty-limit assumption (\ref{dirty-1}), we expand the action in the gradients of $Q$ [simultaneously expanding the logarithm in Eq.\ (\ref{action}) in $\delta\mathbf h$] and obtain, in the usual manner,\cite{Efetov,AST} the action (\ref{S0-S1-S2}) with
\begin{gather}
\mathcal S_0 = \pi\nu \int d^3 \mathbf r \STr\left\{ \frac D4 (\nabla Q)^2 + \left[ i \hat\tau_3 (E - \hat{\boldsymbol\sigma} \mathbf h) - \hat\Delta \right] Q \right\} , \label{S_0} \\
\mathcal S_1 = -i\pi\nu \int d^3\mathbf r\; \delta\mathbf h (\mathbf r) \STr\bigl( \hat\tau_3 \hat{\boldsymbol\sigma} Q(\mathbf r) \bigr), \label{Sh}
\end{gather}
and
\begin{multline} \label{S2}
\mathcal S_2 = \frac 12 \STr \int d^3 \mathbf r d^3 \mathbf r' \frac{d^3 \mathbf p d^3 \mathbf p'}{(2\pi)^6} \left[ T \hat\tau_3 \hat{\boldsymbol\sigma} \delta\mathbf h T^{-1} \right]_{\mathbf r} \frac{e^{i\mathbf p (\mathbf r -\mathbf r')}}{\xi -\frac{i\Lambda}{2\tau}} \\
\times \left[ T \hat\tau_3 \hat{\boldsymbol\sigma} \delta\mathbf h T^{-1} \right]_{\mathbf r'} \frac{e^{-i\mathbf p' (\mathbf r -\mathbf r')}} {\xi' -\frac{i\Lambda}{2\tau}}\, ,
\end{multline}
where the local matrix $T(\mathbf r)$ parametrizes rotations of the $Q$ matrix,
\begin{equation}
 Q = T^{-1} \Lambda T \, , \qquad \Lambda = \hat\tau_3 \, ,
\end{equation}
and $\xi=\mathbf p^2/(2m)-\mu$, $\xi'={\mathbf p'}^2/(2m)-\mu$.

The integrals over $\mathbf p$ and $\mathbf p'$ in Eq.\ (\ref{S2}) may be computed, giving rise to the kernel decaying at the elastic scattering length $l$. Assuming that the $Q$ matrix changes on length scales much longer than $l$, we can put $T(\mathbf r)= T(\mathbf r')$ in Eq.\ (\ref{S2}) and arrive at
\begin{multline} \label{Shh}
\mathcal S_{2} = - \frac{\pi^2 \nu^2}2 \int d^3 \mathbf r d^3 \mathbf r' \; \frac{\sin^2 \left(k_F |\mathbf r-\mathbf r'| \right)}{\left(k_F |\mathbf r-\mathbf r'| \right)^2}\, e^{-|\mathbf r-\mathbf r'|/l} \\
\times \delta h_i(\mathbf r) \delta h_j(\mathbf r') \STr \bigl( \hat\tau_3 \hat\sigma_i Q(\mathbf r) \hat\tau_3 \hat\sigma_j Q(\mathbf r) \bigr)\, .
\end{multline}

\subsection{Local spin-flip term}

Averaging $\mathcal S_{2}$ over magnetic disorder (\ref{disorder}) produces the spin-flip term (\ref{action-sf}) with
\begin{equation} \label{gamma-local}
\Gamma_\mathrm{sf}^{ij} = \pi \nu \int d^3\mathbf r\; F_{ij}(r) \frac{\sin^2 (k_F r)}{(k_F r)^2} e^{-r/l} \, .
\end{equation}

Note that this expression generalizes the result of Abrikosov and Gor'kov for pointlike impurities.\cite{AG} In that work, the magnetic disorder was assumed delta-correlated, which corresponds to  $F_{ij}(\mathbf r -\mathbf r') = \delta_{ij}\delta(\mathbf r -\mathbf r')/(6\pi\nu\tau_s)$ and $\Gamma^{ij}_\mathrm{sf} = \delta_{ij}/(6\tau_s)$. Our derivation extends that result to the medium-ranged disorder with correlation lengths up to $l$. As we shall see below, the contribution (\ref{gamma-local}) is dominant as long as $a \ll l$, and therefore in this regime we can neglect the factor $e^{-r/l}$ and arrive at Eq.\ (\ref{sf-1-2}).

\subsection{Nonlocal spin-flip term}

Averaging $\mathcal S_1$ over the magnetic disorder (\ref{disorder}) produces the contribution to the action
\begin{multline} \label{S1}
-\frac 12 \bigl< (\mathcal S_1)^2 \bigr>= \frac{\pi^2 \nu^2}2 \int d^3 \mathbf r d^3 \mathbf r' F_{ij}(\mathbf r-\mathbf r') \\
\times \STr \bigl( \hat\tau_3 \hat\sigma_i Q(\mathbf r) \bigr) \STr \bigl( \hat\tau_3 \hat\sigma_j Q(\mathbf r') \bigr)\, .
\end{multline}
While the main part of the action $\mathcal S_0$ contains only one $\STr$, this contribution is a product of two supertraces. We assume that the saddle point $Q_0$ is supersymmetric (or replica symmetric), and then the contribution of Eq.\ (\ref{S1}) vanishes at such a saddle point. However, taking into account non-supersymmetric (non-replica-symmetric) fluctuations around the saddle point produces a non-negligible contribution containing only one $\STr$.

In order to average Eq.\ (\ref{S1}) over fluctuations of $Q$, we parametrize those fluctuations by local rotation matrices $W$, anticommuting with $Q_0$:
\begin{equation} \label{Q}
Q = Q_0 + i Q_0 W + \dots
\end{equation}
The effective action for $W$, extracted from the $\mathcal S_0$ part, to the Gaussian order is
\begin{equation} \label{S0}
\mathcal S_W = \frac{\pi\nu D}4 \int d^3 \mathbf r \STr (\nabla W)^2\, .
\end{equation}
Note that since Eq.\ (\ref{S1}) involves correlations of $W$ at the length scale of order $a$, we only need to take into account short-wavelength fluctuations of $W$. Therefore, we neglect the terms containing $E$, $\mathbf h$, $\Delta$, and $\nabla Q_0$ in Eq.\ (\ref{S0}), as well as the self-consistent ``screening'' by the effective spin flip $\Gamma_\mathrm{sf}$ [which produces the infrared cutoff for the action (\ref{S0}); see also Sec.~\ref{subsection:screening} below], under the assumption
\begin{equation} \label{assumption-nonlocal}
E_\mathrm{Th}, E, \Delta, h, \Gamma_\mathrm{sf} \ll E_a \, .
\end{equation}

To average over the fluctuations with the action (\ref{S0}) for $W$ anticommuting with $Q_0$,
we employ the following contraction rule:\cite{Efetov,Mirlin}
\begin{multline} \label{reconnecting}
\bigl< \STr(A_1 W(\mathbf r)) \STr(A_2 W(\mathbf r')) \bigr>_{\mathcal S_W} \\
= \frac 1{\pi\nu D} [\nabla^{-2}]_{\mathbf r \mathbf r'} \STr(A_1 Q_0 A_2 Q_0 -A_1 A_2)
\end{multline}
for any operators $A_1$ and $A_2$ [here
$[\nabla^{-2}]_{\mathbf r \mathbf r'}= - (4\pi |\mathbf r - \mathbf r'|)^{-1}$
is the kernel of the inverse Laplacian in three dimensions].
Applying this identity to averaging the contribution
(\ref{S1}) with the matrix $Q$ parametrized by Eq.\ (\ref{Q}), one
finds that $-\frac 12 \left< (\mathcal S_1)^2 \right>_{\mathcal S_W}$
is given by the usual spin-flip term (\ref{action-sf}) with
\begin{equation} \label{gamma-nonlocal}
\Gamma_\mathrm{sf}^{ij} = -\frac 1D \int d^3 \mathbf r\; F_{ij}(\mathbf r) [\nabla^{-2}]_{\mathbf 0 \mathbf r} \, ,
\end{equation}
which coincides with Eq.\ (\ref{sf-3}).

\subsection{Discussion of the two contributions}

While the two contributions (\ref{two-contributions}) come from two different terms in the sigma-model action, we shall see that they, in fact, correspond to the two limiting cases of magnetic disorder, as depicted in Fig.~\ref{fig:diagr}.

First of all, let us compare the magnitude of the two spin-flip rates. The ``local'' spin-flip rate (\ref{gamma-local}) has the order of magnitude
\begin{align}
\Gamma_\mathrm{sf}^\mathrm{(local)} & \sim \nu \left( \delta h \right)^2 a^3\, , \qquad
a \ll k_F^{-1}\, , \\
\Gamma_\mathrm{sf}^\mathrm{(local)} & \sim \nu \left( \delta h \right)^2 a k_F^{-2} \, , \qquad
k_F^{-1} \ll a \ll l\, , \\
\Gamma_\mathrm{sf}^\mathrm{(local)} & \sim \nu \left( \delta h \right)^2 l k_F^{-2} \, , \qquad
l \ll a\, .
\end{align}
On the other hand, the ``nonlocal'' contribution (\ref{gamma-nonlocal}) can be estimated as
\begin{equation}
\Gamma_\mathrm{sf}^\mathrm{(nonlocal)} \sim \frac{(\delta h)^2 a^2}{D} \sim \nu (\delta h)^2 \frac{a^2}l k_F^{-2} \, .
\end{equation}
Therefore, the nonlocal term (\ref{gamma-nonlocal}) dominates for $a \gg l$, while the local term (\ref{gamma-local}) becomes dominant at $a \ll l$ [strictly speaking, the nonlocal term is only defined for $a\gg l$, see our discussion below].

\begin{figure}
 \includegraphics[width=0.95\hsize]{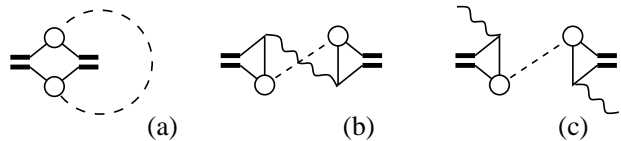}
\caption{Diagrammatic representations of the (a)~local and (b)~nonlocal
contributions in the sigma-model calculation. Thick double lines correspond to
the $Q$ matrix; closed loops of thin solid lines indicate the $\STr$ operation;
open circles denote the ``magnetic vertices''
$\hat\tau_3\hat{\boldsymbol\sigma}$. Dashed lines are the magnetic disorder
correlation functions $F_{ij}$; wavy lines are the propagators of the $W$ field.
Panel (c) shows a potentially dangerous contribution to the propagator of the
$W$ field. To avoid divergencies, a self-consistent ``screening'' of the $W$
propagator needs to be taken into account.}
 \label{fig:diags-2}
\end{figure}

Second, we can graphically represent the two contributions as shown in
Fig.~\ref{fig:diags-2} [panels (a) and (b)]. Correlation functions of
the $Q$ matrices in the sigma model correspond to the diffusion
ladders in the conventional diagrammatic technique,\cite{Efetov} and
therefore the sigma-model diagrams represented in
Figs.~\ref{fig:diags-2}(a) and~\ref{fig:diags-2}(b) translate to the
processes shown in Fig.~\ref{fig:diagr}. Thus, while formally our
derivation produces the sum of the two terms
$\Gamma_\mathrm{sf}^\mathrm{(local)} +
\Gamma_\mathrm{sf}^\mathrm{(nonlocal)}$, only the local term should be
kept in the regime $a \ll l$, and only the nonlocal term --- in the
opposite regime $a \gg l$. At the same time, the nonlocal term has
been derived under an implicit assumption $a \gg l$, since in
Eq.\ (\ref{reconnecting}) we have used the diffusion propagator for the
correlation function of $W(\mathbf r)$ and $W(\mathbf r')$ at
distances of order $a$. Therefore, at the intermediate length scales
$a \sim l$, none of the terms (\ref{gamma-local}) or
(\ref{gamma-nonlocal}), nor their sum, provide an accurate result for
the spin-flip rate. To calculate an effective spin-flip rate at
$a \sim l$, a crossover from the ballistic to the diffusive motion needs
to be taken into account. This calculation is performed in
Sec.~\ref{section:diagram} in the diagrammatic language (a similar
calculation in the context of a collinear periodic magnetization has
been done in Refs.~\onlinecite{Buzdin_av1,Buzdin_av2}).

Finally, we would like to comment on the applicability conditions of our derivation. For the derivation of the local term (at $a\ll l$), we only need the conditions (\ref{dirty-1}) and (\ref{dirty-2}). For the nonlocal term (at $a \gg l$), we have also assumed the condition (\ref{assumption-nonlocal}). Altogether, the applicability conditions may be reformulated in the universal form (\ref{hierarchy-general}).

Note that the propagator of the $W$ field (used in the calculation of the nonlocal term) gets, in principle, renormalized by higher-order contributions. One of the potentially dangerous corrections is shown in Fig.~\ref{fig:diags-2}(c). To avoid infrared divergencies in this correction, one needs to take into account that the diffusive propagator (\ref{S0}) for the $W$ field is cut off at large distances by a certain screening length. This screening is discussed in more detail (in the language of the Usadel equations) in Sec.~\ref{subsection:screening}; the resulting screening length is given by Eq.\ (\ref{rstar-screening}). If this screening is taken into account, then higher-order corrections to the propagator of the $W$ field may be neglected.

\section{Derivation from the Usadel equations}
\label{section:usadel}

In this section, we present an alternative derivation of the spin-flip term in the regime of long-range correlations ($a\gg l$) by directly averaging the Usadel equations\cite{Usadel,LO,RS,BVE_review} over the magnetic disorder, following an approach similar to that of Ref.~\onlinecite{IF}. In this way, we derive Eqs.\ (\ref{Usadel_symm}) and (\ref{sf-3}) and lift the assumptions of ``self-averaging'' and of ``being away from the gap edge'' imposed in Ref.~\onlinecite{IF}.

In the regime of long-range correlations, the general assumption (\ref{hierarchy-general}) may be simplified as
\begin{equation} \label{hierarchy}
E_\mathrm{Th}, E, \Delta, h, \delta h \ll E_a
\end{equation}
[since $\Gamma_\mathrm{sf}\sim (\delta h)^2/E_a$, as we derive below].

\subsection{Effective spin-flip rate}

We start from the Usadel equation containing the exchange field,\cite{BVE_review,IF}
\begin{equation} \label{Us0}
D\nabla \left( \check G \nabla \check G \right) +\left[ i E \hat\tau_3 \hat\sigma_0 - \hat\Delta \hat\sigma_0 - i\hat\tau_3 (\mathbf H \hat{\boldsymbol\sigma}), \check G \right] =0,
\end{equation}
where the gradient term can also be rewritten as
\begin{equation} \label{grad}
\nabla \left( \check G \nabla \check G \right) = \frac 12 \left[ \check G, \nabla^2 \check G \right]
\end{equation}
due to the normalization condition $\check G^2 =1$. Here, $\mathbf H$ is the total realization-dependent exchange field, containing a smooth background field $\mathbf h$ and a Gaussian disorder $\delta\mathbf h$ obeying Eq.\ (\ref{disorder}):
\begin{equation}
\mathbf H = \mathbf h + \delta\mathbf h .
\end{equation}
The exact solution of Eq.\ (\ref{Us0}) can be written as the sum
\begin{equation}
\check G = \check g + \delta \check g
\end{equation}
of the disorder-averaged part $\check g = \langle \check G \rangle$ and the $\delta\check g$ part that depends on the realization of the magnetic disorder and averages to zero.

As confirmed by our further derivation, under the assumption
(\ref{hierarchy}), the realization-dependent part $\delta\check g$ is
small, $|\delta\check g| \ll 1$, and it is sufficient to consider it
linear in $\delta\mathbf h$. Our aim is to obtain the equation for
$\check g$, the disorder-averaged Green function.
Note that $\check g$
is not simply the zeroth order over $\delta\mathbf h$: under our
assumptions it is also influenced by the averages containing the second
order over $\delta\mathbf h$.

Averaging the normalization condition $\check G^2 = 1$ over the magnetic disorder and neglecting the $\langle \delta\check g^2 \rangle$ term [this is possible under the assumption (\ref{hierarchy}), see Sec.~\ref{subsection:screening} for details], we obtain the normalization condition $\check g^2 =1$. Then the realization-dependent part must obey the relation
\begin{equation} \label{gdg}
\left\{ \check g, \delta\check g \right\} =0.
\end{equation}

Averaging Eq.\ (\ref{Us0}) over the magnetic disorder and taking into account Eq.\ (\ref{grad}), we find
\begin{multline} \label{Uss1}
\frac D2 \left[ \check g, \nabla^2 \check g \right] +\left[ i E \hat\tau_3
\hat\sigma_0 - \hat\Delta \hat\sigma_0 - i\hat\tau_3 (\mathbf h
\hat{\boldsymbol\sigma}), \check g \right] \\
- i \left[ \hat\tau_3 \hat\sigma_i, \left< \delta h_i
\delta\check g \right> \right] =0,
\end{multline}
where the summation over the repeating indices is assumed. Here, we have dropped out the full derivative term containing $\nabla \left< [ \delta\check g, \nabla \delta\check g ] \right>$, since this term has an additional smallness [as confirmed by the result (\ref{linear-perturbation_0}) below].

To calculate the averages in Eq.\ (\ref{Uss1}), we extract from Eq.\ (\ref{Us0}) [we also take into account Eq.\ (\ref{grad})] the linear part in $\delta\mathbf h$:
\begin{multline} \label{linear-perturbation}
\frac D2 \left[ \check g, \nabla^2 \delta\check g \right]
-\left[ \frac{D}{2} \nabla^2 \check g -
i E \hat\tau_3 \hat\sigma_0 + \hat\Delta \hat\sigma_0 +
i\hat\tau_3 (\mathbf h \hat{\boldsymbol\sigma}), \delta\check g \right] \\
= i \left[ \hat\tau_3 (\delta\mathbf h \hat{\boldsymbol\sigma}),
\check g \right] .
\end{multline}
This is a linear equation with respect to $\delta\check g$, with the
source term containing the disorder $\delta\mathbf h$. In order to
find $\delta\check g$ from this equation, we note that the first term
is the largest one, since the derivatives apply to the
fast function $\delta\check g$ which follows $\delta\mathbf h$ and
hence changes on the scale of $a$. The second term in the left-hand side is
smaller, according to our assumption (\ref{hierarchy}), and for most
purposes we may neglect it. Employing Eq.\ (\ref{gdg}),
we then obtain
\begin{equation} \label{linear-perturbation_0}
\delta\check g = \frac iD \left( \nabla^{-2} \delta h_i \right) \check g \left[ \hat\tau_3 \hat\sigma_i, \check g \right].
\end{equation}

Now the disorder-induced part of Eq.\ (\ref{Uss1}) after simple algebraic manipulations takes the standard form of the spin-flip term:
\begin{equation}
 - i \left[ \hat\tau_3 \hat\sigma_i, \left< \delta h_i
\delta\check g \right> \right] = -\Gamma_\mathrm{sf}^{ij}
\left[ \hat\tau_3 \hat\sigma_i \check g \hat\tau_3 \hat\sigma_j,
\check g \right] ,
\end{equation}
where we have defined
\begin{equation}
\Gamma_\mathrm{sf}^{ij} = -\frac 1D \bigl< \delta h_i \nabla^{-2} \delta h_j \bigr>
\end{equation}
[which is equivalent to Eq.\ (\ref{sf-3})]. Thus, we finally arrive at Eq.\ (\ref{Usadel_symm}).

\subsection{Self-consistent screening of disorder-induced correlations}
\label{subsection:screening}

In the above calculation, we have neglected the terms containing the
disorder averages $\langle \delta \check g^2 \rangle$, at the same
time keeping the terms with
$\langle \delta h_i \delta\check g \rangle$.
While the exact calculation of the neglected terms appears
to be a delicate problem, we can estimate their order of magnitude
(and thus justify neglecting them) from simple arguments.

To estimate those averages, we should express $\delta\check g$ from
Eq.\ (\ref{linear-perturbation}) and then average over $\delta\mathbf
h$. However, in this procedure, we cannot limit ourselves to the
approximation (\ref{linear-perturbation_0}), which would produce a
divergence:
\begin{equation} \label{no-screening}
\bigl< \delta\check g^2 \bigr> \sim \frac 1{D^2} \int\frac{F_{ij}(p)}{p^4} d^3 \mathbf p \quad \to \quad \infty .
\end{equation}

To regularize this infrared divergence, one needs to take into account
the second term in Eq.\
(\ref{linear-perturbation}), which provides an effective cut-off for
the integral (\ref{no-screening}). The corresponding ``screening
length'' $R^*$ is determined by the largest of the energy scales
$E_\mathrm{Th}$, $E$, $\Delta$, $h$ [corresponding to the second
term in Eq.\ (\ref{linear-perturbation})], and
$\Gamma_\mathrm{sf}$ [the latter energy scale appears if one
self-consistently includes the spin-flip terms in Eq.\
(\ref{Usadel_symm}) in our expansion]:
\begin{equation} \label{rstar-screening}
R^* \sim \sqrt{\frac D{\max(E_\mathrm{Th}, E, \Delta, h, \Gamma_\mathrm{sf})}}\, .
\end{equation}
As a result, the neglected terms may be estimated as
\begin{equation}
\bigl< \delta\check g^2 \bigr> \lesssim \frac{F(p=0)}{D^2} R^* \sim \left(\frac{\delta h}{E_a} \right)^2 \left( \frac{R^*}{a} \right)\, .
\label{screening-estimate}
\end{equation}

For our approximation (neglecting those terms), we require that they are much smaller than one, and that they produce corrections smaller than the spin-flip term $\Gamma_\mathrm{sf} \sim (\delta h)^2 / E_a$. The first condition translates into the requirement $\delta h \ll E_a$ (under this condition, the neglected terms are small). The second condition gives the additional constraint $E_\mathrm{Th}, E, \Delta, h \ll E_a$ (under this condition, the spin-flip term is the main effect of the disorder). Altogether, the conditions of applicability of our derivation can now be formulated as Eq.\ (\ref{hierarchy}).

Note that taking into account the cut-off length $R^*$ allows us to get rid of the assumption of the ``self-averaging disorder'' made in Ref.~\onlinecite{IF} to guarantee the convergence of the integral (\ref{no-screening}). Now we see that the condition (\ref{hierarchy}) is sufficient for that.

The above treatment of the self-consistent screening of disorder-induced
correlations has been performed under the assumption of a three-dimensional
magnetic disorder. At lower dimensions (e.g., if the magnetic disorder
forms layers), a nonlocal interference becomes important, and replacing
the effect of magnetic disorder by a local spin flip may not always be
possible.\cite{IF} Technically, the three-dimensionality is used in
our derivation of the estimate (\ref{screening-estimate}). If one repeats
the calculation in a dimension lower than three, one finds that the
smallness of $\langle \delta\check g^2 \rangle$ cannot be guaranteed.
So in low dimensions our approximation breaks down, and we expect
that the role of magnetic disorder becomes in this case nonlocal
and non-universal.

\subsection{Effective spin-flip scattering at the edge of the minigap}

In the above derivation, we assumed that the linear operator acting on
$\delta\check g$ in Eq.\ (\ref{linear-perturbation}) is
invertible [and that we can keep only the $\nabla^{-2}$ term in its
inverse, which led us to Eq.\ (\ref{linear-perturbation_0})]. As it
was pointed out in Ref.~\onlinecite{IF}, this assumption breaks down
at the edge of the minigap, where the solution of the Usadel equation
bifurcates. In that case, the linear-order perturbation theory over
$\delta\mathbf h$ breaks down, which is formally reflected in the
non-invertibility of the linear operator in Eq.\ (\ref{linear-perturbation}).

However, the non-invertibility of the operator (and the corresponding zero mode) is associated with the boundary conditions at the edge of the system and, hence, with the length scale $L$ of the system size. On the other hand, the spin-flip processes producing the scattering rate (\ref{sf-3}) and corresponding to Fig.~\ref{fig:diagr}(b) are associated with the length scale $a$ (with $a\ll L$) and do not depend on the boundary conditions. Therefore we expect that the same form of the spin-flip term remains valid also near the minigap edge.

The problem with the perturbative expansion at the minigap edge is due to the
fact that the minigap itself depends on the spin-flip
rate,\cite{Crouzy-minigap} which leads to a non-analytic shift of the
saddle point of the action (\ref{action-full}) as a function of the
disorder $\delta\mathbf{h}$. A correct way to include the effect of
disorder in the vicinity of the minigap is to first calculate the
spin-flip rate from Eq.\ (\ref{sf-3}) [this expression is independent
of the particular saddle-point solution $\check g ( \mathbf r )$]
and then to recalculate the new saddle-point solution (with a new
value of the minigap) with this spin-flip rate using the Usadel
equations. We thus conclude that our result for the
effective spin flip remains valid at all energies across the minigap
edge, and the assumption of Ref.~\onlinecite{IF} about the
invertibility of the operator in the left-hand side of
Eq.\ (\ref{linear-perturbation}) is unnecessary.

\section{Diagrammatic representation}
\label{section:diagram}

An alternative method to obtain the spin-flip scattering rate involves a direct
calculation of the diagrams shown in Fig.~\ref{fig:diagr}. These diagrams are
superconducting cooperon propagators: they are built of the retarded Green
function for particles [with dispersion $\xi(\mathbf{p})$] and the retarded
Green function for holes [with dispersion $-\xi(\mathbf{p})$]. The latter can be
converted into the advanced function for particles at the opposite energy, which we shall use below.
These cooperon soft modes naturally arise in the diagrammatic expansion of
the non-linear sigma model discussed in Sec.~\ref{section:sigma-model}. Thus the
diagrams of Fig.~\ref{fig:diagr} directly correspond to the sigma-model diagrams
of Fig.~\ref{fig:diags-2}.

The cooperon modes are massless in the absence of superconducting correlations (i.e., become singular if both the total external momentum and energy are zero). Once the spin flip is taken into account, the cooperon acquires a mass which is directly related to the spin-flip rate.

Superconducting correlations in the system (e.g., due to the proximity effect) also
produce a mass for the cooperon. We shall neglect this effect in the calculation
of the spin-flip rate due to the conditions (\ref{hierarchy-general}). All the
mechanisms resulting in a mass of the cooperon, including the spin flip, are weak
and their contributions may be calculated independently.

In order to calculate the effective spin-flip rate, we shall evaluate the two
diagrams depicted in Fig.~\ref{fig:diagr} at the zero total momentum and zero
energy. The diagrams (a) and (b) yield the following contributions to the
cooperon self energy:
\begin{multline} \label{C-a}
 \gamma_{(a)}^{ij}
  = \int \frac{d^3 {\mathbf q}}{(2\pi)^3}\, F_{ij}(q) \\
    \times \int \frac{d^3 {\mathbf p}}{(2\pi)^3}\,
    G^R(\mathbf{p} + \mathbf{q}) G^A(-\mathbf{p} - \mathbf{q})
    G^R(\mathbf{p}) G^A(-\mathbf{p})
\end{multline}
and
\begin{multline} \label{C-b}
 \gamma_{(b)}^{ij}
  = 2 \int \frac{d^3 {\mathbf q}}{(2\pi)^3}\, F_{ij}(q) C(q) \\
    \times \left| \int \frac{d^3 {\mathbf p}}{(2\pi)^3}\,
      G^R(\mathbf{p} + \mathbf{q}) G^R(\mathbf{p}) G^A(-\mathbf{p})
    \right|^2\, ,
\end{multline}
respectively, where $G^{R,A}(\mathbf{p}) = [-\xi(\mathbf p) \pm \frac i{2\tau}]^{-1}$
are the zero-energy retarded and advanced Green functions. $C(q)$ denotes the
cooperon containing only the potential impurities,
\begin{equation}
 C(q)
  = \frac{1 + B(q) + B^2(q) + \ldots}{2 \pi \nu \tau}
  = \frac{1}{2 \pi \nu \tau}\, \frac{1}{1 - B(q)}\, ,
\end{equation}
where
\begin{equation}
 \label{B}
 B(q)
  = \frac{1}{2 \pi \nu \tau} \int \frac{d^3 {\mathbf p}}{(2\pi)^3}\,
    G^R(\mathbf{p} + \mathbf{q}) G^A(-\mathbf{p})
\end{equation}
is a single ladder rung containing a disorder line and two Green functions. The
factor of 2 in Eq.\ (\ref{C-b}) comes from two possible diagrams of type (b).

The total spin-flip scattering rate is the sum of Eqs.\ (\ref{C-a}) and
(\ref{C-b}) with a coefficient that can be easily determined from
comparing to the limit of pointlike magnetic impurities.\cite{AG}
In that limit, $F_{ij}(q)$ is actually independent of the momentum $q$ and only
the term (\ref{C-a}) contributes, yielding
\begin{equation}
 \gamma_{(a)}^{ij}
  = 4\pi^2\nu^2\tau^2 \int d^3 {\mathbf r}\, F_{ij}(r)\, .
\end{equation}
By comparing with Eq.\ (\ref{sf-1}), we arrive at the
result for the spin-flip rate:
\begin{equation} \label{sf-tot}
 \Gamma_\mathrm{sf}^{ij}
  = \frac 1{4\pi\nu\tau^2} \left(
      \gamma_{(a)}^{ij} + \gamma_{(b)}^{ij}
    \right)\, .
\end{equation}

Now the calculation of the spin-flip rate can be conveniently performed
in the two overlapping regimes: the medium-to-long-range magnetic
correlations ($a \gg k_F^{-1}$) and the short-to-medium-range magnetic
correlations ($a \ll l$).

In the medium-to-long-range regime, the integrals in Eqs.\ (\ref{C-a}) and
(\ref{C-b}) are restricted to $q \ll k_F$ and can be performed by
using the integration over $\xi$ in the vicinity of the Fermi surface.
First, we use the identity
\begin{equation} \label{unitarity}
 G^R(\mathbf{p}) G^A(-\mathbf{p})
  = i \tau \left[ G^R(\mathbf{p}) - G^A(-\mathbf{p}) \right]\,
\end{equation}
and discard all the integrals containing only
retarded or only advanced Green functions (since
they have all the poles lying in the same half-plane of the variable $\xi$).
This allows us to re-express Eqs.\ (\ref{C-a}) and (\ref{C-b})
in terms of the function $B(q)$ defined in Eq.~(\ref{B}):
\begin{gather}
 \label{sf-loc}
 \gamma_{(a)}^{ij}
  = 4\pi\nu\tau^3 \int \frac{d^3 {\mathbf q}}{(2\pi)^3}\, F_{ij}(q) B(q)\, ,\\
 \label{sf-nonloc}
 \gamma_{(b)}^{ij}
  = 4\pi\nu\tau^3 \int \frac{d^3 {\mathbf q}}{(2\pi)^3}\,
    \frac{F_{ij}(q) B^2(q)}{1 - B(q)}\, .
\end{gather}
As a result, Eq.\ (\ref{sf-tot}) leads to
\begin{equation} \label{sf-tot-long}
\Gamma_\mathrm{sf}^{ij}
  = \tau \int \frac{d^3 {\mathbf q}}{(2\pi)^3}\,
    \frac{F_{ij}(q) B(q)}{1 - B(q)}\, .
\end{equation}
An explicit calculation of $B(q)$ (by using an integration over $\xi$)
gives
\begin{multline} \label{B-explicit-1}
 B(q)
  = \frac{1}{2 \pi \nu \tau}
    \int \frac{d^3 {\mathbf p}/(2 \pi)^3}{
      (-\xi - \mathbf{v q} + \frac{i}{2\tau}) (-\xi - \frac{i}{2\tau})
    } \\
  = \left< \frac{1}{1 + i \tau \mathbf{v q}} \right>_\mathbf{v} =
\frac{\arctan(q l)}{q l}
\end{multline}
(the averaging here is over the Fermi surface). Substituting this expression into Eq.\
(\ref{sf-tot-long}), we obtain the result
(\ref{sf-2-3}) for the medium- and long-range-correlation regimes.

In the short-to-medium-range regime, the integrals (\ref{C-a}) and (\ref{C-b})
are determined by the momenta $q \gg l^{-1}$, which allows us to
neglect the contribution (\ref{C-b}) in favor of (\ref{C-a}). If the correlation
length of the magnetic disorder becomes comparable to $k_F^{-1}$, the integrals of $G^R G^R$
and $G^A G^A$ cannot be neglected any more, and the integral (\ref{C-a})
should be calculated in a different way.
The main contribution to the $\mathbf{p}$ integral comes
from the intersection of the two mass shells of the ``width'' $l^{-1}$ shifted
by the vector $\mathbf{q}$. Using the inequality $q \gg l^{-1}$, we
approximate $G^R (\mathbf{p}) G^A (-\mathbf{p})$
by the delta-function and obtain
\begin{multline} \label{B-explicit-2}
 \int \frac{d^3 {\mathbf p}}{(2\pi)^3}\,
    G^R(\mathbf{p} + \mathbf{q}) G^A(-\mathbf{p} - \mathbf{q})
    G^R(\mathbf{p}) G^A(-\mathbf{p}) \\
  = 4 \pi^2 \tau^2 \int \frac{d^3 {\mathbf p}}{(2 \pi)^3}\,
    \delta[\xi(\mathbf{p})]\, \delta[\xi(\mathbf{p} + \mathbf{q})] \\
  = \frac{2 \pi^2 \nu \tau^3}{q l}\, \theta(2 k_F - q)\, .
\end{multline}
Substituting this expression into Eq.~(\ref{C-a}), we arrive at
the short-to-medium-range crossover result (\ref{sf-1-2}).

Of course, in the quasiclassical limit $k_F l \gg 1$ considered in this paper,
one is allowed to combine the two overlapping regimes into
a single formula
\begin{equation} \label{sf-general}
 \Gamma_\mathrm{sf}^{ij}
  = \tau \int\limits_{q < 2 k_F} \frac{d^3 {\mathbf q}}{(2\pi)^3}\,
    \frac{F_{ij}(q) \arctan(q l)}{q l - \arctan(q l)}\, ,
\end{equation}
which reproduces both the short-to-medium- and medium-to-long-range crossover
results.

\section{Conclusions}
\label{section:conclusions}

To summarize, we have analyzed the effect of magnetic inhomogeneities in disordered superconducting systems. Our approach covers magnetic disorder of various correlation lengths, and thus extends the theory of Abrikosov and Gor'kov for magnetic impurities,\cite{AG} as well as some earlier studies of superconductivity in systems with inhomogeneous magnetism.\cite{IF,Buzdin_av1,Buzdin_av2} The main conclusion of our work is that if the correlation length of the magnetic disorder is much shorter than all the macroscopic scales in the problem [condition (\ref{hierarchy-general})], then the effect of the magnetic disorder may be incorporated as an effective local spin-flip rate (in the same form as for magnetic impurities\cite{AG}). We have obtained exact expressions for the effective spin-flip rate, under the assumption of a Gaussian magnetic disorder.

While the exact expressions for the effective spin-flip rate are
probably of mainly academic interest, we believe that our results will
be helpful for estimating the spin-flip effects induced by
inhomogeneities in various experimental setups. As an example of such
an application, we consider the experiments on SFS $\pi$-junctions,
where the spin flip plays an important role\cite{Ryazanov1,Ryazanov2}
(it manifests itself in the difference between the length scales
involved in the decay and oscillations of the critical current as a
function of the junction thickness). If we apply our estimates to the
experimental data reported in Ref.~\onlinecite{Ryazanov2} (assuming
$\delta h \sim h$), then we arrive at the estimate of the disorder
correlation scale $a\sim 2$~nm. Note that this correlation scale is of
the order of the length scales associated with the uniform component
$h$ of the magnetic field and with the resulting spin-flip rate, thus,
this example is at the borderline of applicability of our theory. The
estimated size of inhomogeneities is apparently too small for domains
(in recent experiments on CuNi films similar to those used in the
$\pi$-junctions, domains of size about 100~nm have been
reported\cite{Vinnikov}). However, our estimates are consistent with
earlier indications of clusters of magnetic Ni atoms in such
alloys\cite{Ni1,Ni2} (inhomogeneities inside the domains).

Finally, we would like to comment on comparison between the effective
spin-flip rates due to two cases of inhomogeneous magnetization:
disordered and periodic ones. The disordered case is considered in the
present paper, while specific realizations of periodic magnetic
structures were studied before in
Refs.~\onlinecite{Buzdin_av1,Buzdin_av2,IF,BVE}. The obtained results
for the spin-flip rate are all of the same order of magnitude,
differing only by numerical factors. This suggests that
there is probably no qualitative difference between the effective
spin-flip rates in disordered and periodic magnetic structures,
as long as the characteristic length scale of inhomogeneities
is sufficiently small.

\begin{acknowledgments}
We are grateful to M.~V.\ Feigel'man, B.\ Crouzy, and S.\ Tollis for helpful discussions. This work was supported by the Swiss National Foundation, the Dynasty Foundation, RF Presidential Grants Nos.\ MK-4421.2007.2 
and NSh-5786.2008.2, RFBR Grants Nos.\ 07-02-01300 and 07-02-00310, and the program ``Quantum physics of condensed matter'' of the RAS. The hospitality of the Institute of Theoretical Physics at EPFL, where the main part of this work was done, is gratefully acknowledged.
\end{acknowledgments}


\begin{thebibliography}{99}
\bibitem{T}
The energy $E$ in the condition (\ref{hierarchy-general}) is the characteristic
energy of electrons in the problem. For finite-temperature
calculations, this energy scale may be determined by temperature.
\bibitem{spin-flip-matrix}
In the most anisotropic case, the spin-flip matrix is described by 6
independent parameters. Depending on the symmetry of the
problem, the number of parameters may be smaller, e.g., in the most
symmetric isotropic case, the spin-flip matrix is diagonal and
described by one scattering time. In the case of a ferromagnetic order, when
disorder is anisotropic (one axis is special), the spin-flip rates
in the directions along the ferromagnetic axis and perpendicular to
it may be different: in this case, two scattering times would be
necessary.\cite{Gorkov-Rusinov,Fulde-Maki,Houzet}
\bibitem{E}
Note that although we can use the Matsubara technique, which is standard in the case of replicas, here we choose the real-energy description.
\end{thebibliography}
\end{document}